# CVGG-Net: Ship Recognition for SAR Images Based on Complex-Valued Convolutional Neural Network

Dandan Zhao, Zhe Zhang, *Member, IEEE,* Dongdong Lu, Jian Kang, *Member, IEEE,* Xiaolan Qiu, *Senior Member, IEEE*, and Yirong Wu

*Abstract*— Ship target recognition is a vital task in synthetic aperture radar (SAR) imaging applications. Although convolutional neural networks have been successfully employed for SAR image target recognition, surpassing traditional algorithms, most existing research concentrates on the amplitude domain and neglects the essential phase information. Furthermore, several complex-valued neural networks utilize average pooling to achieve full complex values, resulting in suboptimal performance. To address these concerns, this paper introduces a Complex-valued Convolutional Neural Network (CVGG-Net) specifically designed for SAR image ship recognition. CVGG-Net effectively leverages both the amplitude and phase information in complex-valued SAR data. Additionally, this study examines the impact of various widely-used complex activation functions on network performance and presents a novel complex max-pooling method, called Complex Area Max-Pooling. Experimental results from two measured SAR datasets demonstrate that the proposed algorithm outperforms conventional real-valued convolutional neural networks. The proposed framework is validated on several SAR datasets.

## I. INTRODUCTION

SYNTHETIC aperture radar (SAR) is a high-resolution active microwave imaging sensor that operates without limitations related to weather or time [1], playing a crucial role in both military and civilian applications.

SAR target recognition involves using variety of methods to identify the category of targets in SAR images, and it represents a popular research direction in microwave remote sensing applications[2].

Presently, SAR target recognition methods can be roughly categorized into traditional methods, deep learning-based methods, and complex information-based methods. Traditional SAR image target recognition methods rely on two key technologies: feature extraction and target recognition. SAR images possess unique geometric, mathematical, and electromagnetic features that significantly differ from optical images. Common feature extraction methods for SAR images include template matching [3] and feature fusion [4]. Once features are extracted, an appropriate classifier is selected to recognize and classify them. Popular classifiers currently include support vector machines [5-6], sparse representation classification[7], and others. However, most traditional recognition methods are adaptations of optical pattern recognition, characterized by complex feature designs and limited recognition rates.

With the advent of deep learning technology, frameworks such as Convolutional Neural Networks (CNNs) and Fully Convolutional Neural Networks (FCNs) have demonstrated promising performance in SAR image target recognition, increasingly becoming mainstream in the field. Chen et al. [8] and Ding et al. [9] employed CNNs for SAR image target recognition, achieving impressive results on the MSTAR dataset. Concurrently, Lin et al. [10] integrated channel convolution and attention mechanisms, effectively focusing on the target area within SAR images. Deep learning-based methods can automatically extract features and perform recognition tasks, eliminating the need for manual feature design or complex physical modeling. However, the majority of deep learning-based work emphasizes amplitude information, while phase information remains a crucial factor in SAR image target recognition [11]. SAR utilizes microwave coherent imaging, rendering SAR images complex-valued [12]. Compared to real values, complex values offer superior representation and generalization characteristics[13]. Consequently, there is an urgent need to develop SAR image target recognition methods that fully harness the advantages of complex-valued information.

In 2017, literature [14] first introduced the CV-CNN, which employs complex average pooling to achieve better results than real-valued networks on multi-channel POLSAR image classification tasks. Yu et al. [15] proposed the CV-FCNN, consisting only of convolutional layers and utilizing 1x1 complex convolution to learn cross-channel feature information, while Zhang et al. [16] presented the CV-MotionNet, a complex-valued convolutional neural network architecture that eliminates the need for motion compensation in classifying SAR moving ship targets. However, recognizing targets from SAR images can be challenging due to the effects of coherent speckle noise [17], which results in the discretization of target pixels and poor target separability. Although complex-valued SAR images contain rich phase information, relying solely on amplitude information makes it difficult to obtain satisfactory recognition results.

To address these issues, we propose a complex-valued convolutional neural network (CVGG-Net) for target recognition in SAR images that fully utilizes the amplitude and phase information in complex-valued SAR data.

Corresponding author: Zhe Zhang (zhangzhe01@aircas.ac.cn).
D. Zhao is with School of Information and Communication Engineering, Hainan University, Haikou 570228, China.
D. Zhao, Z. Zhang, D. Lu, X. Qiu and Y. Wu are with Suzhou Key Laboratory of Microwave Imaging, Processing and Application Technology, and Suzhou Aerospace Information Research Institute, Suzhou, China.
D. Zhao, Z. Zhang, D. Lu, X. Qiu and Y. Wu are also with Aerospace Information Research Institute, Chinese Academy of Sciences, Beijing, China.
J. Kang is with Soochow University, Suzhou, China.

Moreover, we introduce a novel complex max-pooling method, termed "Complex Area Max-Pooling", which aids the network in extracting more effective features. Compared to traditional real-valued CNNs, the proposed method achieves higher accuracy.

## II. PROPOSED TARGET RECOGNITION METHOD

In recent years, CNNs have made significant strides in computer vision tasks. The CVGG-Net, proposed in this study, is based on the architecture of VGG[18] and incorporates the complex-valued convolution operation from [19]. Specifically designed for recognizing target objects in SAR images, the network accepts single-channel input. As illustrated in Fig. 1, the proposed CVGG-Net comprises 13 complex-valued convolutional blocks, 5 complex area max-pooling layers, 3 complex fully connected layers, 1 amplitude evaluation layer, and 1 softmax layer.

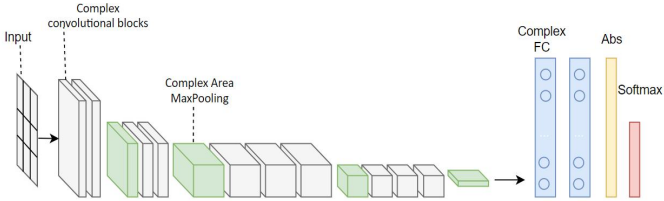

**Fig. 1.** Architecture of the CVGG-Net.

CVGG-Net is designed with a VGG-like architecture, closely resembling VGG16. All layers within the network are complex-valued, allowing complex convolutional layers to extract information present in the amplitude and phase of complex SAR images. This approach enables the extraction of richer features compared to traditional deep learning methods. Complex area max-pooling is utilized to reduce the number of network parameters and enhance overall robustness. The complex fully connected layers further extract target features, which are then converted to real values in the last layer for calculating cross-entropy loss with labels. Finally, the network leverages a softmax layer for target recognition.

### A. Complex-valued convolutional blocks

A complex-valued convolutional block consists of a complex convolutional layer, a complex batch normalization layer, and a complex activation function layer, as shown in Fig. 2.

*1) Complex Convolutional Layer:* Complex convolution is an extension of traditional convolution in computer vision that operates in the complex domain. Unlike traditional convolution, which only utilizes amplitude information, complex convolution extracts target features by using both amplitude and phase information present in complex-valued SAR images. This approach results in complex-valued operations being added to the traditional convolution process. Experimental results have demonstrated that complex convolutional layers outperform conventional convolutional layers, indicating their superiority in extracting meaningful target features.

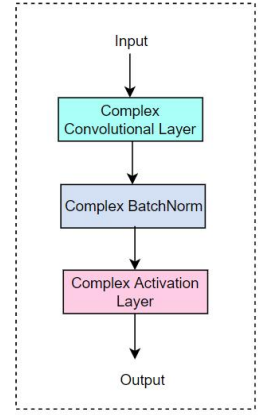

**Fig. 2.** Schematic diagram of complex-valued convolutional blocks.

According to [19], when the convolution operation is extended to the complex field, the complex vector $I = x + yj$ and the complex convolution kernel $W = A + Bj$ perform corresponding element-wise multiplication and summation operations, where $j = \sqrt{-1}$. Separating the real and imaginary values in the complex feature layer, the complex-valued convolution is equivalent to:

$$W * I = (A + Bj) * (x + yj)$$
$$= (A * x - B * y) + (A * y + B * x)j \quad (1)$$

In the right hand side of the above equation, * stands for traditional (real) convolution. $x$ and $y$ represent the real and imaginary parts of the complex-valued vector respectively, $A$ and $B$ represent the real and imaginary parts of the complex convolution kernel, respectively. It can be seen that one complex-valued convolution operation is equivalent to four conventional convolution operations, as shown in Fig. 3.

*2) Complex Batch Normalization Layer:* In deep neural networks, batch normalization is often employed to stabilize the intermediate output values of each layer, promote model convergence, and mitigate the risk of overfitting [20].

Batch normalization of complex values can be scaled by the square root of the variance of the two principal components, real and imaginary, through dividing the zero-centered data $(X - E(X))$ by the square root of the $2 \times 2$ covariance matrix $V$ [19].

$$\tilde{X} = \frac{X - E(X)}{\sqrt{V}} \quad (4)$$

$$V = \begin{pmatrix} cov(\Re\{x\}, \Re\{x\}) & cov(\Re\{x\}, \Im\{x\}) \\ cov(\Im\{x\}, \Re\{x\}) & cov(\Im\{x\}, \Im\{x\}) \end{pmatrix} \quad (5)$$

*3) Complex Activation Function Layer:* In order to handle complex-valued representations, a family of complex activation functions have been proposed. CRelu introduced in [21], extending the traditional Relu activation function to the complex domain.

$$CRelu(z) = Relu(\Re\{z\}) + jRelu(\Im\{z\}) \quad (6)$$

CRelu applies separate Relu activations to both the real and imaginary parts of neurons, satisfying the Cauchy Riemann equation when both parts are either positive or negative.

Existing literature lacks a definitive consensus on the most suitable activation function for complex-valued neural networks. The primary requirement for the function is to be nonlinear, with gradients that do not suffer from issues such as explosion or vanishing during training [22]. In this paper, we have evaluated several common complex-valued activation functions from the perspective of ease of implementation. After extensive experimentation, we have selected CReLU as the optimal activation function.

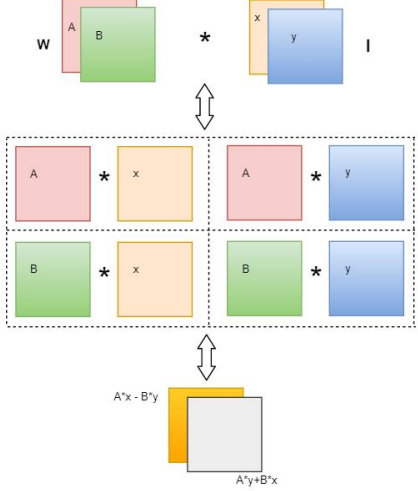

**Fig. 3.** Schematic diagram of complex-valued convolution structure.

*B. Complex Area Max-Pooling*

As shown in Fig. 4(a), many researchers split complex-valued data into imaginary and real parts, and perform real value-based complex max-pooling on the real and imaginary parts, respectively. However, this approach is unreasonable and unsuitable for complex-valued SAR data. Fig. 4(b) illustrates the amplitude-based complex max-pooling (CAMaxPool) proposed in [23], which preserves the complex values corresponding to the coordinates with the maximum amplitude. Building on CAMaxPool, this paper further proposes complex area-based max-pooling, which selects the coordinates corresponding to the maximum area, as shown in Fig. 4(c).

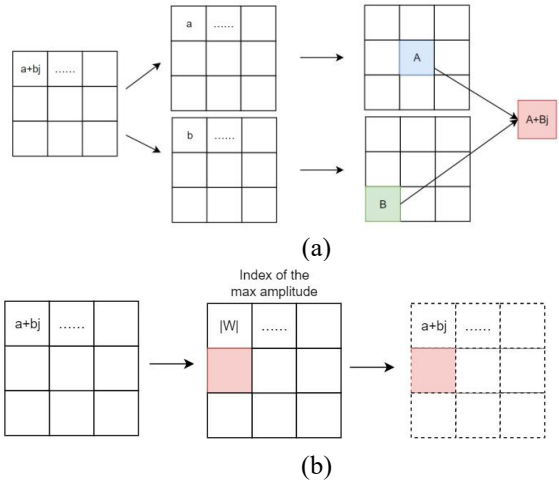

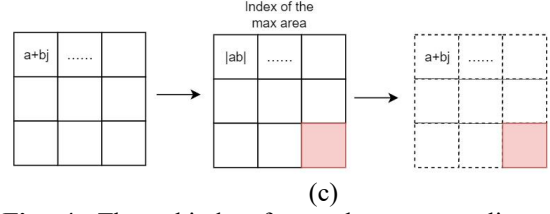

**Fig. 4.** Three kinds of complex max-pooling. (a) Real value-based complex max-pooling. (b) Amplitude-based complex max-pooling. (c) Area-based complex max-pooling.

These two complex-valued max-pooling methods can be viewed as a choice between the values of $f_1$ and $f_2$. As shown in Fig. 5, $f_1$ represents the length of the vector OA, and $f_2$ is the area of the triangle AOB.

$$f_1(z) = \sqrt{x^2 + y^2} \quad (7)$$
$$f_2(z) = |xy| \quad (8)$$

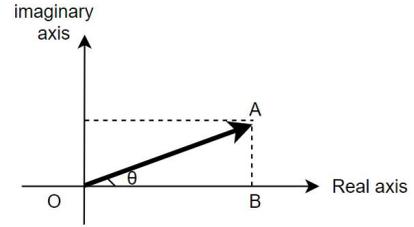

**Fig. 5.** Diagram of coordinates.

The proposed complex area max-pooling introduces an alternative option for selecting pooling coordinates. In essence, the proposed complex area max-pooling tends to retain elements where both real and imaginary parts are present. Although in the SAR context, we cannot provide an appropriate physical explanation for this pooling strategy, our experiments have demonstrated that area-based max-pooling outperforms canonical amplitude-based complex max-pooling.

## III. EXPERIMENTAL RESULTS

*A. Experimental data and platform*

Two SAR datasets are utilized to validate the proposed framework. One is CSRSDD (Complex SAR images Rotation Ship Detection dataset) annotated in [24]. The other is OpenSARship, a public dataset released by Shanghai JiaoTong University in 2017 [25].

*1) CSRSDD.* All data were collected in GF-3 Spotlight (SL) mode with 1m resolution and HH or HV polarization mode. Each image size was 1024x1024 pixels, containing more abundant features. The rotating box is used to mark the target, and the annotation format refers to the DOTA dataset format [26]. The data is primarily concentrated in the port area, with the offshore scene accounting for more than 80%, featuring a complex background and significant interference. To suit the recognition task, we cut the targets out according to the annotation files. The amplitude slices are saved as .tiff files, and the complex slices as .mat files, with the largest dimensions being 610x944 pixels and the smallest 12x34 pixels, as shown in Fig. 6. Five categories are screened out, as presented in Table I.

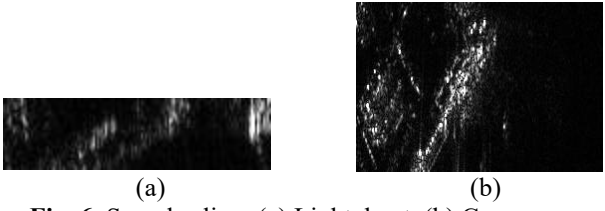

**Fig. 6.** Sample slices.(a) Light_boat. (b) Cargo.

TABLE I
THE NUMBER OF SAMPLES OF TARGETS ON CSRSDD

| Class | cargo | ship1 | ship2 | light_boat | other | All |
|---|---|---|---|---|---|---|
| Train | 67 | 99 | 464 | 533 | 145 | 1308 |
| Test | 28 | 42 | 199 | 228 | 62 | 559 |

*2) OpenSARship.* The dataset was collected from Sentinel-1 images and contains ground range multi-look products and slant range single-look complex products with both VV and VH polarization. In this paper, three types of ship targets—Bulk carrier, Cargo, and Tanker—are selected for research. Ship categories with too few samples are discarded, and each type of ship is randomly divided into training and testing sets according to Table II.

TABLE II
THE NUMBER OF SAMPLES OF TARGETS ON OPENSARSHIP

| Class | Bulk carrier | Container ship | Tanker | All |
|---|---|---|---|---|
| Train | 169 | 169 | 169 | 507 |
| Test | 164 | 404 | 73 | 641 |

In order to facilitate network processing, all image slices in both CSRSDD and OpenSARship are normalized, filled or cropped to 224x224 pixels.

All experiments in this paper were conducted on Pytorch 1.8.1 and a device with Ubuntu18.04 system. Hardware features include Quadro RTX 8000 GPU (40GB memory), CPU AMD 3950x and 64G RAM. Due to the limitation of GPU memory, we set the batch size to be 32, the initial learning rate as 0.0001, and adopt Adam optimizer for optimization, train it for 100 epochs in total.

### B. Comparison of Proposed network and Real-valued Networks

As comparison for benchmark, a small complex-valued network CVnet5 with five complex-valued convolution blocks is built in this paper. Its structure is shown in Fig. 7.

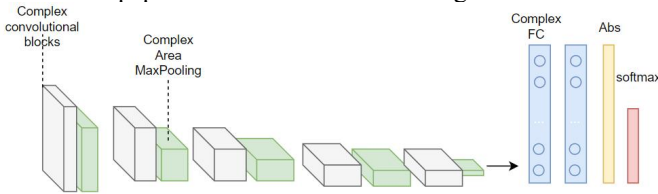

**Fig. 7.** Architecture of the CVnet5.

To verify the effectiveness of the proposed method, we compare it with commonly used real-valued networks such as ResNet18, VGG16, and Net5, which is the real-valued network corresponding to CVnet5. Each experiment is performed 10 times, and the average value is taken as the final experimental result. The results on the CSRSDD dataset and OpenSARship dataset are shown in Table III.

TABLE III
COMPARISON BETWEEN DIFFERENT METHODS ON CSRSDD AND OPENSARSHIP

| | Vnet5 | Resnet18 | Vgg16 | CVGG-Net |
|---|---|---|---|---|
| CSRSDD | 73.47 | 75.57 | 75.01 | **79.57** |
| OpenSARship | 69.73 | 70.75 | 70.05 | **71.07** |

On the CSRSDD dataset, Vnet5, which is the corresponding real-valued network of CVnet5, has the least number of layers and the weakest feature extraction ability, achieving only a 73.47% accuracy. In comparison, ResNet18 and VGG16 achieve 75.57% and 75.01% accuracy, respectively. Our proposed CVGG-Net achieves 79.57% accuracy. On the OpenSARship dataset, Vnet5, ResNet18, and VGG16 achieve recognition rates of 69.73%, 70.75%, and 70.05%, respectively, while our proposed CVGG-Net achieves 71.07%. It can be seen that our proposed CVGG-Net outperforms the real-valued networks.

### C. Comparison of Different Complex Activation Functions

CRelu, CTanh, CElu, and CPrelu are extensions of the real-valued activation functions ReLU, Tanh, Elu, and Prelu, respectively. The recognition rates of these four activation functions on the CSRSDD and OpenSARship datasets are tested using Cvnet5, as illustrated in Fig. 7. The recognition rates are presented in Table IV.

TABLE IV
COMPARISON BETWEEN DIFFERENT COMPLEX ACTIVATION FUNCTIONS ON CSRSDD AND OPENSARSHIP

| | CRelu | CTanh | CElu | CPrelu |
|---|---|---|---|---|
| CSRSDD | **77.92** | 60.11 | 75.29 | 77.41 |
| OpenSARship | **70.59** | 58.42 | 68.49 | 69.27 |

It can be seen from Table III that CTanh has the worst performance, CRelu has a better effect on complex network. Therefore, in the following experiments, CRelu is used for activation functions of complex network.

### D. Comparison Between Complex Amplitude-based max-pooling and Complex Area-based max-pooling

In this paper, the amplitude-based complex max-pooling CAMaxPool proposed in [23] and the complex area max-pooling proposed by us is compared, respectively, applied to complex-valued networks Vnet5 and CVGG-Net, and the experimental results on CSRSDD dataset and OpenSARship dataset are obtained in Table V.

TABLE V
COMPARISON BETWEEN OUR METHOD AND OTHER COMPLEX VALUED NETWORKS

| | CVnet5 + amplitude-based max-pooling | CVnet5 | CVGG-Net + amplitude-based max-pooling | CVGG-Net |
|---|---|---|---|---|
| CSRSDD | 77.17 | 77.92 | 78.86 | **79.57** |
| OpenSARship | 70.59 | 70.89 | 70.20 | **71.07** |

It can be observed that, on the CSRSDD dataset, CVnet5 and CVGG-Net with amplitude-based max-pooling achieve

77.17% and 78.86% recognition rates, respectively. With our area-based max-pooling, the network performance is improved by 0.75% and 0.71%, respectively. On the OpenSARship dataset, the network performance is improved by 0.3% and 0.87%, respectively, with area-based max-pooling. These results indicate that the proposed area-based max-pooling is more effective than amplitude-based max-pooling for complex-valued networks.

*D. Comparison with other complex-valued networks*

To verify the effectiveness of the proposed method, we compared the recognition results with other complex-valued convolutional networks, including CVnet5 and the complex-valued network CV-Net proposed in [17]. As shown in Table VII, the performance of the proposed method is superior to other methods on both the CSRSDD and OpenSARship datasets. Even when CV-Net is combined with our complex area max-pooling, the recognition rates are increased by 0.07% and 0.23%. This improvement can be attributed to the fact that the CVGG-Net, based on complex-valued convolution blocks and complex area max-pooling, is better suited for SAR image target recognition tasks.

TABLE VII
COMPARISON BETWEEN OUR METHODS AND OTHER COMPLEX-VALUED NETWORKS

|  | CVnet5 | CVGG-Net | CV-Net[17] | CV-Net[17] + area-based max-pooling |
|---|---|---|---|---|
| CSRSDD | 77.92 | **79.57** | 75.72 | 75.85 |
| OpenSARship | 70.89 | **71.07** | 66.70 | 66.93 |

## Ⅳ. CONCLUSION

Traditional target recognition methods for SAR images often neglect the phase information, which is crucial for recognition accuracy. Considering that SAR images are inherently complex-valued, we propose a complex-valued convolutional neural network method named CVGG-Net for target recognition in SAR images. Additionally, we introduce a novel complex max-pooling method based on area, which outperforms other complex max-pooling methods. Our experimental results, obtained from CSRSDD and OpenSARship datasets, demonstrate the effectiveness of our proposed method, including the complex area max-pooling and CVGG-Net algorithms. In the future, we plan to extend complex-valued convolutional neural networks to other fields, such as SAR object detection and semantic segmentation.